\documentstyle[12pt,iopfts,psfig]{ioplppt}

\begin{document}

\title{SEARCH FOR PRODUCTION OF STRANGELETS IN QUARK MATTER
USING PARTICLE CORRELATIONS}
\author{
        S.~Soff, D.~Ardouin\footnote
{on leave from University of Nantes, U.M.R. Subatech}, 
        C.~Spieles, S.~A.~Bass, H.~St\"ocker}
        
\address{
        Institut f\"ur Theoretische Physik, J. W. Goethe-Universit\"at,\\
        Postfach 11 19 32, D-60054 Frankfurt am Main, Germany;
\footnote{supported by GSI, BMBF, DFG and Buchmann Fellowship}}

\author{D.~Gourio, S.~Schramm}
\address{
        GSI Darmstadt, Postfach 11 05 52, D-64220 Darmstadt, Germany;}

\author{C.~Greiner}
\address{Institut f\"ur Theoretische Physik, J. Liebig-Universit\"at,\\ 
        Heinrich-Buff-Ring 16, D-35392 Gie{\ss}en, Germany;}

\author{R.~Lednicky}
\address{Institute of Physics of the Academy of Sciences 
of the Czech Republic,\\
        Na Slovance 2, 18040 Prague 8, Czech Republic;}

\author{V.~L.~Lyuboshitz}
\address{JINR Dubna, PO Box 79, Moscow, Russia;}

\author{J.-P.~Coffin, C.~Kuhn}
\address{CRN Strasbourg, Universit\'e L. Pasteur, Strasbourg, France.}

\begin{abstract}
We present a new technique for observing the strangelet production in 
quark matter based on unlike particle correlations. A simulation is 
presented with a two-phase thermodynamical model.
\end{abstract}

%%%%%%%%%%%%%%%%%%%%%%%%%%%%%%%%%%%%%%%%%%%%%%%%%%%%%%%%%%%%%%%%%%%%
\section{Motivation}
%%%%%%%%%%%%%%%%%%%%%%%%%%%%%%%%%%%%%%%%%%%%%%%%%%%%%%%%%%%%%%%%%%%%

It has been speculated that the observation of strange quark matter 
droplets might be an unambiguous way to follow up
the transient existence of a quark gluon plasma (QGP) 
expected to be created in ultrarelativistic 
heavy ion collisions. Due to the Pauli
principle, replacing several up and down quarks (u,d) by strange
quarks (s), will lower the Fermi energy and the mass of strangelets consisting of
multi-quark droplets. Their mass can thus become smaller than
the mass of a corresponding ordinary nucleus (with the same baryon number $B$) and
may then correspond to a stable state \cite{Bodmer71,Witten84}. 
The mass per baryon of a strangelet may also be intermediate between
the $\Lambda$ mass and the nucleon mass thus producing a metastable state
which then cannot decay into $\Lambda$'s \cite{ChinKerman79}. 
Both predictions are found in the MIT
bag model, when its energy density constant $B^{1/4}$ 
is varied ($B$ is the crucial parameter).  
Large $B^{1/4}$ values ($>\,180 \,{\rm MeV}$) lead to large quark matter 
droplet masses and therefore to instability versus strong decays. 
Strangelets are found to be stable for $B^{1/4} \le 150\,{\rm  MeV}$ and 
metastable for intermediate $B^{1/4}$ values \cite{Barz88,CGreiner88}.

Several experiments are carried out at the Brookhaven AGS (E864, E878) and the  
CERN SPS (NA52), which look for strangelet production (small  $Z/A$ ratios) \cite{NA52,E864,E878}. 
Here we present a novel method, which can allow for the measurement of the transient 
strange quark matter state, even if it decays on strong interaction time scales. 
The method is based on unequal particle correlations; they are sensitive to the time 
dependent charge and space-time expansion of the system.

%%%%%%%%%%%%%%%%%%%%%%%%%%%%%%%%%%%%%%%%%%%%%%%%%%%%%%%%%%%%%%%%%%%%
\section{The Two-Phase Thermodynamical Model}
%%%%%%%%%%%%%%%%%%%%%%%%%%%%%%%%%%%%%%%%%%%%%%%%%%%%%%%%%%%%%%%%%%%%

The  dynamical evolution of the mixed phase consisting of a quark 
gluon plasma and hadronic gas can be described in a 
two-phase model which takes into account equilibrium 
as well as non-equilibrium features \cite{CGreiner91,Spieles96}. 

Two main assumptions are made :\\
1. The QGP is surrounded by a layer of hadron gas and 
equilibrium is maintained during the evolution. 

The Gibbs equilibrium conditions write:
\begin{eqnarray}
  P_{\rm qgp}&=&P_{\rm hadr}\nonumber \\
  T_{\rm qgp}&=&T_{\rm  hadr}\nonumber \\
  \mu_{\rm q,qgp}&=& \mu_{\rm q,hadr}\nonumber \\
  \mu_{\rm s,qgp}&=& \mu_{\rm s,hadr} \, .
\end{eqnarray}
2. Non-equilibrium evaporation is incorporated by a time
dependent emission of hadrons from the surface of
the hadronic fireball. The rate of frozen-out hadrons is assumed to be governed 
by the actual hadronic time dependent densities with 
an universal effective relative rate $\Gamma$ \cite{CGreiner91}. 
The volume change $({\rm d}V)$ during the evolution of the 
plasma is determined by the
thermodynamical energy according to:
\begin{equation}
{\rm d}E = T\, {\rm d}S - p\,{\rm d}V + \sum_i \mu_i\,{\rm d} N_i \, .
\end{equation}
In this model the volume increase of the expanding system
competes with the volume decrease caused by freeze-out emission. 
Within this model, it is possible to follow the 
evolution of mass, entropy and strangeness fraction of  
the system. From this, relative yields of particles can be extracted 
which compare reasonably well with experimental data 
\cite{Spieles97}. 

Fig.1 shows the predicted evolution of the quark 
chemical potential $\mu_{\rm q}$. It decreases from $\mu_{\rm q}\approx 115\,{\rm MeV}$ 
to $\mu_q^{\rm final}\approx 15\,{\rm MeV}$  
during the hadronization process. Conversely, the s quark
chemical potential $\mu_{\rm s}$ increases from $\mu_{\rm s}=0\,{\rm MeV}$ 
to several tens of MeV. 
Consequently, the strangeness fraction $f_s=(N_s-N_{\bar{s}})/A_{\rm tot}$ 
increases. This leads to a strange multi-quark object if the bag 
constant $B^{1/4}$ is sufficiently small to allow for cooling of the system (see fig.2).
 
An important new aspect of these calculations is the fact that 
the system necessarily leaves the usual $\mu_{\rm q}-T$ plane of the phase diagram. 
It enters quickly into the strangeness sector. 
Two important features emerge from these calculations \cite{CGreiner87} 
(see fig.3 as an example):\\
1) Strange and anti-strange quarks, produced in equal amount in the
hot plasma, do not hadronize at the same time (in a baryon rich 
system). This so-called "distillation process" \cite{CGreiner87} 
results in the formation of stable or metastable blobs of 
strange matter in the case of low bag model constants. (In this case, a cooling of 
the system is predicted.)
The evaporation from the surface of the hadron gas which is rich of anti-strangeness 
carries away $K^+$ etc. thus charging up the remaining mixed system with net 
positive strangeness.\\
2) The large $f_s$ $(\approx 1.5-2)$ attained 
imply negative charge-to-mass ratios 
$(Z/A=-0.1$ for stable, $Z/A=-0.45$ for metastable droplets).

In the present work, these time 
separation and negative charge values characterize the
transient existence of the plasma. They can be used to search for 
the distillation process by correlations of unlike particles.

%%%%%%%%%%%%%%%%%%%%%%%%%%%%%%%%%%%%%%%%%%%%%%%%%%%%%%%%%%%%%%%%%%%
\section{Influence of Coulomb Interactions on Particle Interferometry}
%%%%%%%%%%%%%%%%%%%%%%%%%%%%%%%%%%%%%%%%%%%%%%%%%%%%%%%%%%%%%%%%%%

In the case of charged particles (like $K^+$, $K^-$), their Coulomb 
interaction modifies the plane and outgoing scattered waves and
the two-particle amplitude takes the form: 
\begin{equation}
\psi_{-\vec{k}}^+={\rm e}^{i\delta}\sqrt{A(\eta)}
\left[{\rm e}^{-i\vec{k}\vec{r}}F(-i\eta,1,i\rho)+\Phi_k^c(r)\right],
\end{equation}
where $\delta$ is the Coulomb s--wave phase shift,
$F(-i\eta,1,i\xi)$ is the confluent hypergeometric function,
$\rho=\vec{k}\vec{r}+kr$, $\eta=(ka)^{-1}$, $|a|$ is the Bohr radius
of the two--particle system, $\Phi_k^c(r)$ is the scattered wave, and
\begin{equation}
A(\eta)=2\pi\eta[\exp(2\pi\eta)-1]^{-1}
\end{equation}  
is the well known Coulomb factor. The $A(\eta)$ factor differs from
unity mainly for $|2\pi\eta|^{-1} < 1$ which means $k < 12$ MeV/c
in the case of the $K^+K^-$ system ($a=-110$ fm).
At high energies 
the two-body Coulomb effect is often treated through a size-independent 
correction factor $A(\eta)$ to the usual  
quantum statistical correlations (i.e., it is assumed that 
$r/|a| \ll 1$ so that $F(-i\eta,1,i\rho)\approx 1$). 
It is thus not used to study the space-time evolution of the system. 
Contrary, at low energies this effect becomes the main tool 
for the study of this evolution due to large transit times. 
In addition, the charged source itself can also affect the above 
two-particle correlation pattern due to the long range interaction 
(e.g. accelerating $K^+$ and deaccelerating $K^-$ for $Z>0$).
This effect is much more clearly visible in the single particle spectra, 
while it nearly cancels out (by taking the denominator) in the correlation function.

Experimental evidence and detailed analysis of such Coulomb two-
and three-body effects on two unlike particle correlation  
functions can be found in \cite{review}. 
The three-body effect itself was observed experimentally more than ten years ago 
\cite{Pochodzalla8586} in intermediate energy heavy ion collisions
and quantitatively analyzed in reactions producing light particles or
unstable fragments (Li$^*$, B$^{*}$) and short-lived resonant states. 
If their decay occurs into two different particles 
(with different charge/mass ratios) within the Coulomb field of the residual 
nuclear system (target/projectile fragments), this field can substantially 
distort their correlation function. 
Thus a complete three-body calculation is required to account for the effect 
of final state interactions. 
One way to look at this width/lifetime effect 
is to compare correlation functions
measured for different  relative velocities \cite{Pochodzalla8586,Pochodzalla87}. 
One then observes a shift 
of the resonance peak which reflects the different Z/A values of the coincident particles 
which suffer the Coulomb accelerations. 
This effect has been later used to explain a variety of two particle 
correlation patterns in the intermediate energy domain \cite{Bertsch93}. 
In the adiabatic limit, it has also been studied 
within a detailed and full three-body quantum approach \cite{Led96a,Martin96}. 
This effect is found to decrease for particles emitted 
at large relative space-time intervals 
or with large momenta with respect to the source.
It is increasing with increasing source charge and also with 
increasing charges and masses of the particles.
 
%%%%%%%%%%%%%%%%%%%%%%%%%%%%%%%%%%%%%%%%%%%%%%%%%%%%%%%%%%%%%%%%%%%%%%%%%
\section{Access to Time Sequence of Emission using Unlike-Particle
Correlations}
%%%%%%%%%%%%%%%%%%%%%%%%%%%%%%%%%%%%%%%%%%%%%%%%%%%%%%%%%%%%%%%%%%%%%%%%%

In addition to the known directional dependence of the
correlations of two identical  particles, it has been shown 
(at low energy heavy ion collisions) 
that the final state interactions between non-identical particles can provide
information about the time-order of emission. 
Thus, based on classical trajectory calculations, 
it was proposed to use the velocity difference spectrum between two 
different charged particles and examine its  
dependence on particle energy in conjunction with the
corresponding relative momentum correlation function \cite{Gel94}. 
Independently, in the quantum approach \cite{Led96a,Led96} 
it was shown that the anisotropy of the
space-time distribution is reflected in the directional dependence of
unlike-particle correlations and can be directly used to measure the 
delays in the emission of particles of different types. 

With the assumption of a sufficiently small density in momentum space, the
correlation is only due to the mutual final state interaction; 
the two-particle amplitude, in the absence of the Coulomb interaction, 
takes the form \cite{Led82}: 
\begin{equation}
\psi_{-\vec{k}}^+={\rm e}^{-i\vec{k}\vec{r}}+\Phi_k(r)\,,
\end{equation}
where $\vec{k}=\vec{q}/2$ and $\vec{r}$ are the relative momentum and spatial 
coordinates of the two particles in their cms system, $\Phi_k(r)$ is
the scattered wave.
%Bethe-Salpeter amplitude of the two-particle state \cite{Led82}. 
The correlation function for two non-identical particles is 
\begin{equation}
1+R(p_1,p_2)=1+\langle|\Phi_k(r)|^2+2\,{\rm Re}\,\Phi_k(r)\cos(\vec{k}\,\vec{r})
-2\,{\rm Im}\,\Phi_k(r)\sin(\vec{k}\,\vec{r})\rangle\,.
\end{equation}
The directional dependence in the correlation function \cite{Led96} 
is contained in the odd term, 
${\rm Im} \,\Phi_k(r) \, {\rm sin}(\vec{k}\,\vec{r})$. 
In the limit of large relative emission times $t=t_1-t_2$, $v|t| \gg r$, 
the vector $\vec{r}$  
is only slightly modified by averaging over the spatial 
location of the emission points in the rest frame of the source. 
So the vector $\vec{r}$ is nearly parallel or antiparallel 
to the pair velocity $\vec{v}$ depending on the sign of the time
difference. The odd part of the correlation function is sensitive to this sign. 
This mean relative emission time, including its sign 
can be determined by comparing the correlation functions $R^+$ and 
$R^-$ \cite{Led96}. 
In these functions the sign of the scalar product $\vec{k}\cdot\vec{v}$ 
is fixed to be larger or smaller than zero (using
their relative angle) respectively. 
Note that for particles of equal masses the sign of 
$\vec{k}\cdot\vec{v}$ is almost equal to the sign 
of the velocity difference $v_1-v_2$. 
This means that the interaction of the two particles will be different in the case where  
the faster particle is emitted earlier as compared     
to the case of its later emission. 

In the case of unlike {\it charged} 
particles, additional odd terms in the 
$\vec{k}\cdot\vec{r}$ product appear due to the 
confluent hypergeometrical function. 
This modifies the plane wave and preserves 
the sensitivity to the time difference 
even in the case of a weak effect of the strong interaction. 
It has also been shown \cite{Led82} that the Coulomb 
field of the residual nucleus slightly affects this result.
 
These predictions have been compared to experimental (p,d) 
correlations measured for the Nb+Pb, Xe+Ti, Xe+Sn collisions
studied at GANIL \cite{Ghi93,Era96,Gou96,Nouais97}. 
The ratio $(1+R^+)/(1+R^-)$ deviates from unity at low $q$. This can be attributed 
to an earlier emission of deuterons rather than protons on the average.  
Thus, it has been experimentally demonstrated for the first time that
the study of the directional asymmetries in the correlations of 
non-identical particles 
provides a unique tool to measure the time-ordering 
of particle emission. This technique will here be extended to 
emission time differences between 
strange, anti-strange and non-strange particles \cite{Led96}. 
We study its sensitivity to the time delays related to the 
production of the transient strange quark matter state.  

%%%%%%%%%%%%%%%%%%%%%%%%%%%%%%%%%%%%%%%%%%%%%%%%%%%%%%%%%%%%%%%%%%%%%%%%%
\section{Calculations with an Event Generator}
%%%%%%%%%%%%%%%%%%%%%%%%%%%%%%%%%%%%%%%%%%%%%%%%%%%%%%%%%%%%%%%%%%%%%%%%%

A quantum two-particle final state interaction code 
\cite{Led82} is used together with a three-body calculation \cite{Led96a} 
to simulate the possible influence on $K^+K^-$ 
correlations caused by 

1) different distributions of the $K^+$ and $K^-$ 
emission times and 

2) by different charge states of the emitting source. 

Results are presented for an event generator 
describing a mid-rapidity source with transverse mass spectra     
conditions typical for AGS experiments. The source is assumed to    
be Gaussian in space coordinates and exponential in time. 
Different sets of time origins are used for the two particles.

Fig.4 shows some simulations with mean emission time differences between 
$K^+$ and $K^-$ of $-5$ to $+10 \,{\rm fm/c}$. 
The figure shows the ratio of the relative momentum correlation functions  
$(1+R^+)/(1+R^-)$ 
as obtained with the above selections for values of the scalar product 
$\vec{k}\cdot\vec{r} >0\,{\rm and}\,< 0$, respectively. 
Clearly, ratios less than one are associated with a scenario where the 
$K^-$'s are emitted later on the average.

Fig.5 shows the dependence of the above ratio on the charge of the source. 
The time difference is fixed to $+10\,{\rm fm/c}$.
The upper diagram shows the result of a $2$-body calculation. 
The two lower diagrams correspond to $3$-body calculations with 
source charges $Z=50$ and $Z=-50$, respectively. We have checked 
that the neglection of the strong final state interaction 
only slightly modifies these results.
Therefore, most of the effect originates from the Coulomb 
interaction between the two kaons with an additional slight 
enhancement (suppression) of this deviation from 
unity created by a positive (negative) source charge. 

%%%%%%%%%%%%%%%%%%%%%%%%%%%%%%%%%%%%%%%%%%%%%%%%%%%%%%%%%%%%%%%%%%%%%%%%%
\section{Calculations with the Two Phase Thermodynamical Model 
coupled to Final State Interactions}
%%%%%%%%%%%%%%%%%%%%%%%%%%%%%%%%%%%%%%%%%%%%%%%%%%%%%%%%%%%%%%%%%%%%%%%%%

The previous section clearly shows the domain of sensitivity of the
$K^+K^-$ correlations to time differences and charge effects. In order to
do similar calculations with a more sophisticated particle source, we 
have described the source evolution using the 
thermodynamical model presented in section 2. Three sets of parameters have been
chosen:\\
{\it set 1 $\&$ 2} (representing Au+Au at AGS energies):\\
initial mass $A= 394$, entropy per baryon $S/A= 10$, initial net
strangeness $f_s= 0$;\\
{\it set 1}: $B^{1/4} = 160\,{\rm MeV}$, strangelets (albeit metastable) are formed.\\
{\it set 2}: $B^{1/4} = 235\,{\rm MeV}$, strangelets are not distilled, their thermal formation 
is also highly suppressed. They are not (meta-)stable.\\
{\it set 3} ( representing S+Au at SPS energies):\\
bag model constant $B^{1/4} = 235\,{\rm MeV}$,\\
initial mass $A= 100$, entropy per baryon $S/A= 45$ , initial net
strangeness $f_s= 0$.\\
Corresponding calculated time evolutions are shown in figs. 6 and 7 for
particle yields, initial quark blob masses and charges. 
In the case of a high bag constant the system heats up slightly. 
The entropy per baryon in the QGP is larger than in the hadronic
phase. This results in a fast hadronization (see the radius and
charge time evolutions). $K^+$ are emitted earlier than $K^-$. 
For low bag constants, a cold strangelet 
emerges in a few tens of fm/c with a radius of $\approx 2.5 \,{\rm fm}$, a 
finite net strangeness content and a {\it \bf negative} charge. 

These predicted time dependent yields are used as input for the final state
interaction calculations. 
Fig.8 shows the calculated correlation function ratios 
$(1+R^+)/(1+R^-)$ 
for 
$K^+K^-$ pairs. 
Similar functional dependencies are found as with the simple event generator
calculations. 

The ratio of the measured correlation functions for 
$\vec{k}\cdot\vec{v}>0$ 
and 
$\vec{k}\cdot\vec{v}<0$ 
shows that the $K^-$ is emitted later in the case of a low bag 
constant value. The effect is seen as a $30-35\%$ dip relative to 
equal mean time production ($R=1$). 
It extends up to $q \approx 30\,{\rm MeV/c}$ for the parameters of {\it set 1}.

A weaker effect ($10\%$) is found, extending up to $q \approx 40\,{\rm MeV/c}$ when 
{\it set 2} (high bag constant) is used. 
The effect is nearly gone for the higher initial entropy {\it set 3}.
This may be attributed to freeze-out time distributions of $K^+$ and
$K^-$ which are quite similar as seen in the predicted yields (figs. 3 and 7). 
The slight difference
between the sets $2 \& 3$ has to be attributed to the   
smaller Coulomb influence of the source itself ($A=100$) in the    
entropy $S/A= 45$ calculation. Our simulations (see fig.5 and 
discussion above ) support this explanation (including the direction of 
this effect) as long as an averaged positive
charge is present (in agreement with the present model calculation displayed 
in fig.6). 

%
%%%%%%%%%%%%%%%%%%%%%%%%%%%%%%%%%%%%%%%%%%%%%%%%%%%%%%%%%%%%%%%%%%%
\section{CONCLUSION AND PERSPECTIVES}

We have presented a novel method, which can allow 
for the measurement of the transient 
strange quark matter state, even if it 
decays on strong interaction time scales. 
The method is based on unequal particle correlations. 
They are sensitive to 
the time dependent charge and space-time expansion of the system. 
The dynamical evolution model for the mixed QGP-hadronic 
phase has been applied to describe 
the strangeness distillation process. 
This predicts a time dependent freeze-out and 
the occurence of low charge-to-mass ratios. 
The existence of strangelets can thus be characterized by the
correlation approach of unlike particles presented here. 
 
Hydrodynamical calculations including the expansion
of the system are currently underway. 
Also microscopic analysis of this effect, based on the UrQMD-simulations, are 
presently under study.  
If combined with the identical particle interferometry (HBT), this  
technique offers the possibility to study different  
expansion processes associated with different transverse motion and 
sizes. 
Correlations of other particles (p$\Lambda$, p$\pi^-$, etc.) are presently 
studied.

%***********************************************************
%\section{Acknowledgements} 
\ack
The authors would like to thank B.~Erazmus, L.~Martin, D.~Nouais and 
C.~Roy for useful discussions and for providing their 
correlation codes.
D.~A. is pleased to thank the Institut f\"ur Theoretische Physik 
at the University of Frankfurt for their invitation and kind 
hospitality.

%***********************************************************
%\newpage
%%%%%%%%%%%%%%%%%%%%%%%%
\section*{References}
%%%%%%%%%%%%%%%%%%%%%%

\Figures

\begin{figure}
\caption{Time evolution of the quark and strange chemical 
potential for parameter {\it set} 3.  
Hadronic and quark matter rapidly moving out of the ($\mu_{\rm q},T$) plane, 
entering the strange sector ($f_s \ne 0$). 
The fit of a static thermal source which yields the same hadron ratios as the 
dynamical model is also depicted.}
\end{figure}

\begin{figure}
\caption{Time evolution of baryon number $A$, strangeness 
fraction $f_s$ and temperature $T$ for 
$S/A_{\rm init}=25$ and $f_s=0.25$ for a (a) high 
($B^{1/4}=235\,{\rm MeV}$) and (b)  
low ($B^{1/4}=145\,{\rm MeV}$) bag constant, respectively.}
\end{figure}

\begin{figure}
\caption{Yields for $\pi^+,\,K^+,\,K^-,\,p$ and $\Lambda$ particles 
as a function of time for the AGS parameter sets $1 \& 2$  
(upper figure $B^{1/4}=160\,{\rm MeV}$, 
lower figure $B^{1/4}=235\,{\rm MeV}$).}
\end{figure}

\begin{figure}
\caption{Simulation of different time origins for $K^+$ and $K^-$ emission with 
the event generator. 
Shown are the ratios $(1+R^+)/(1+R^-)$ as explained in the text.} 
\end{figure}

\begin{figure}
\caption{Simulation of the influence of the source charge state on the 
ratio $(1+R^+)/(1+R^-)$ with the event generator (see text).}
\end{figure}

\begin{figure}
\caption{Radius and charge of the remaining 
quark matter blob as a function of time $t$ 
as calculated with the two-phase thermodynamical model.}
\end{figure}

\begin{figure}
\caption{Particle yields as a function of time $t$ for a bag constant 
$B^{1/4}=235\,{\rm MeV}$, initial entropy 
$S/A=45$, initial baryon number $A_{\rm init}=100$, 
representing an SPS parameter set (set $3$ in the text).}
\end{figure}

\begin{figure}
\caption{Ratios of correlation functions $(1+R^+)/(1+R^-)$ 
for three different parameter sets (see text).}
\end{figure}

\newpage
\clearpage

\samepage{
\psfig{figure=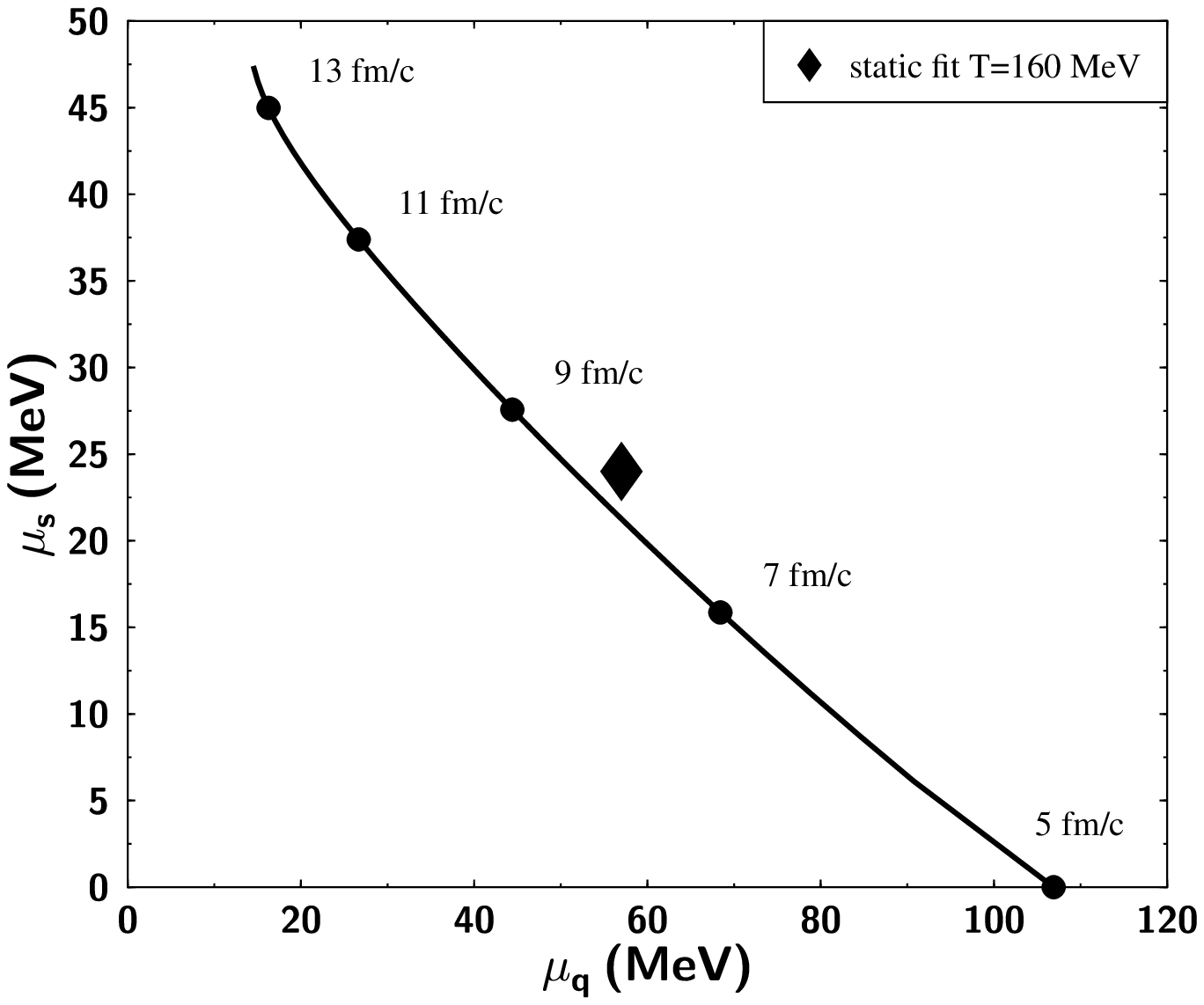}
\vspace*{3cm}
\center{\large \bf Fig.1}}
\newpage
\psfig{figure=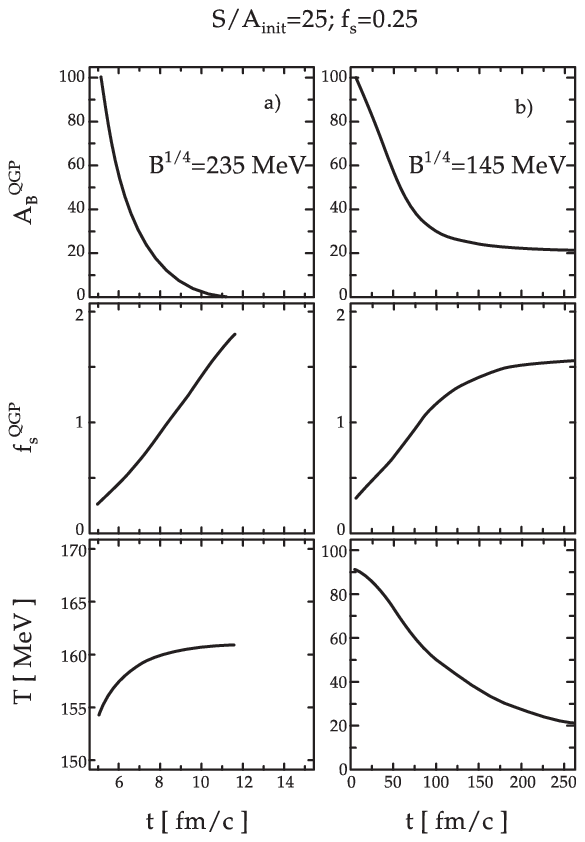,height=12cm,width=8cm}
\vspace*{3cm}
\center{\large \bf Fig.2}
\newpage
\samepage{
\psfig{figure=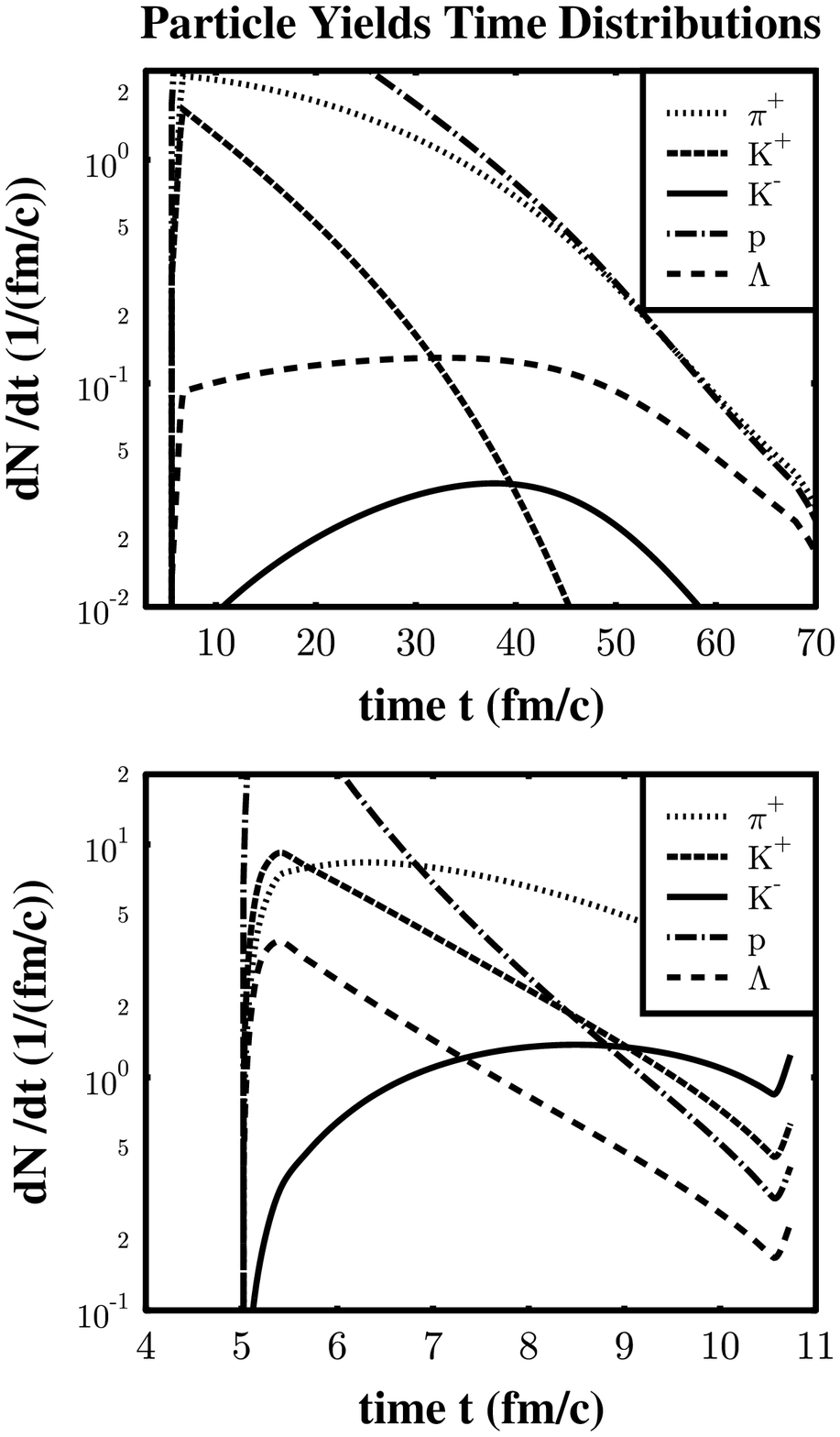}
\vspace*{1cm}
\center{\large \bf Fig.3}}
\newpage
\samepage{\psfig{figure=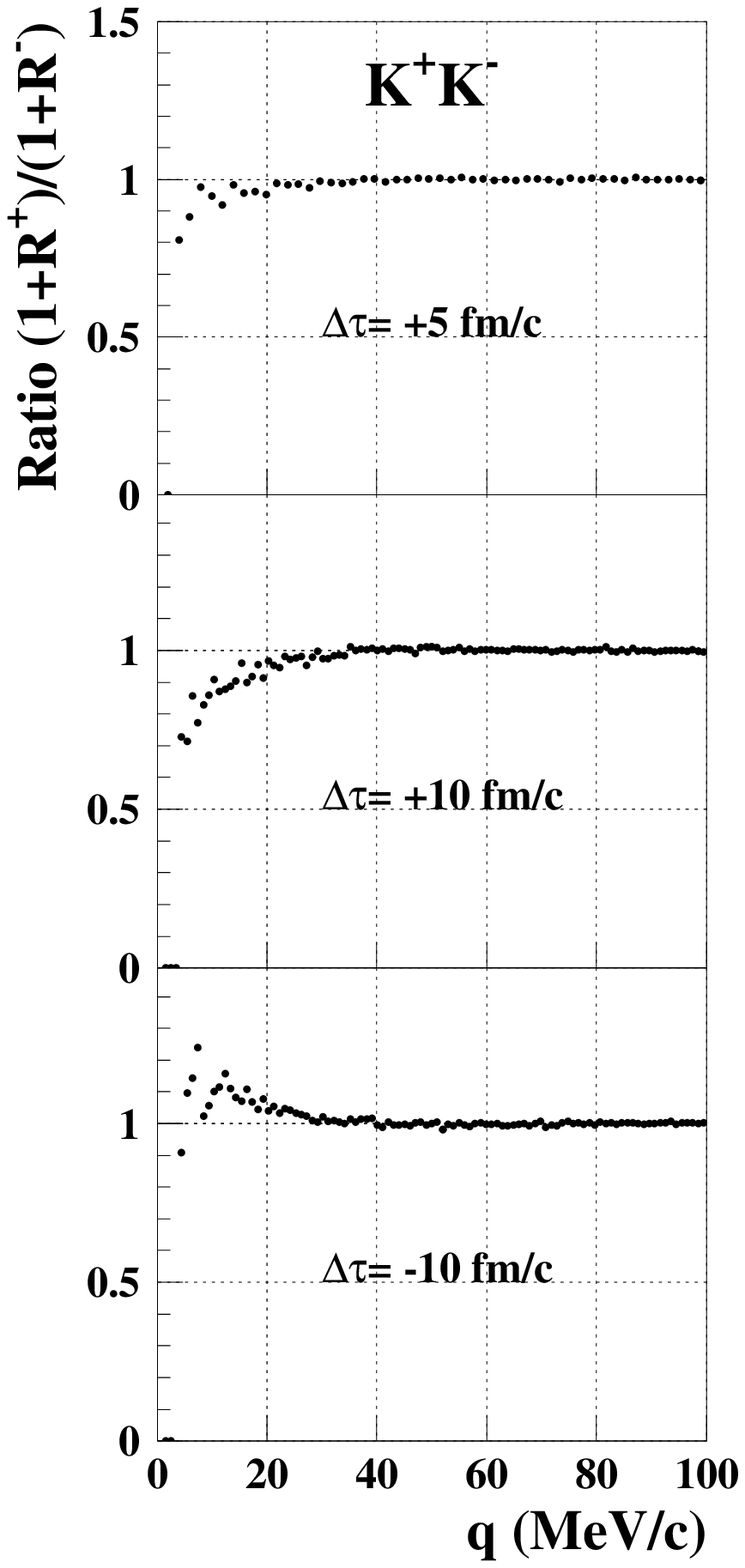}
\vspace*{1cm}
\center{\large \bf Fig.4}}
\newpage
\samepage{
\psfig{figure=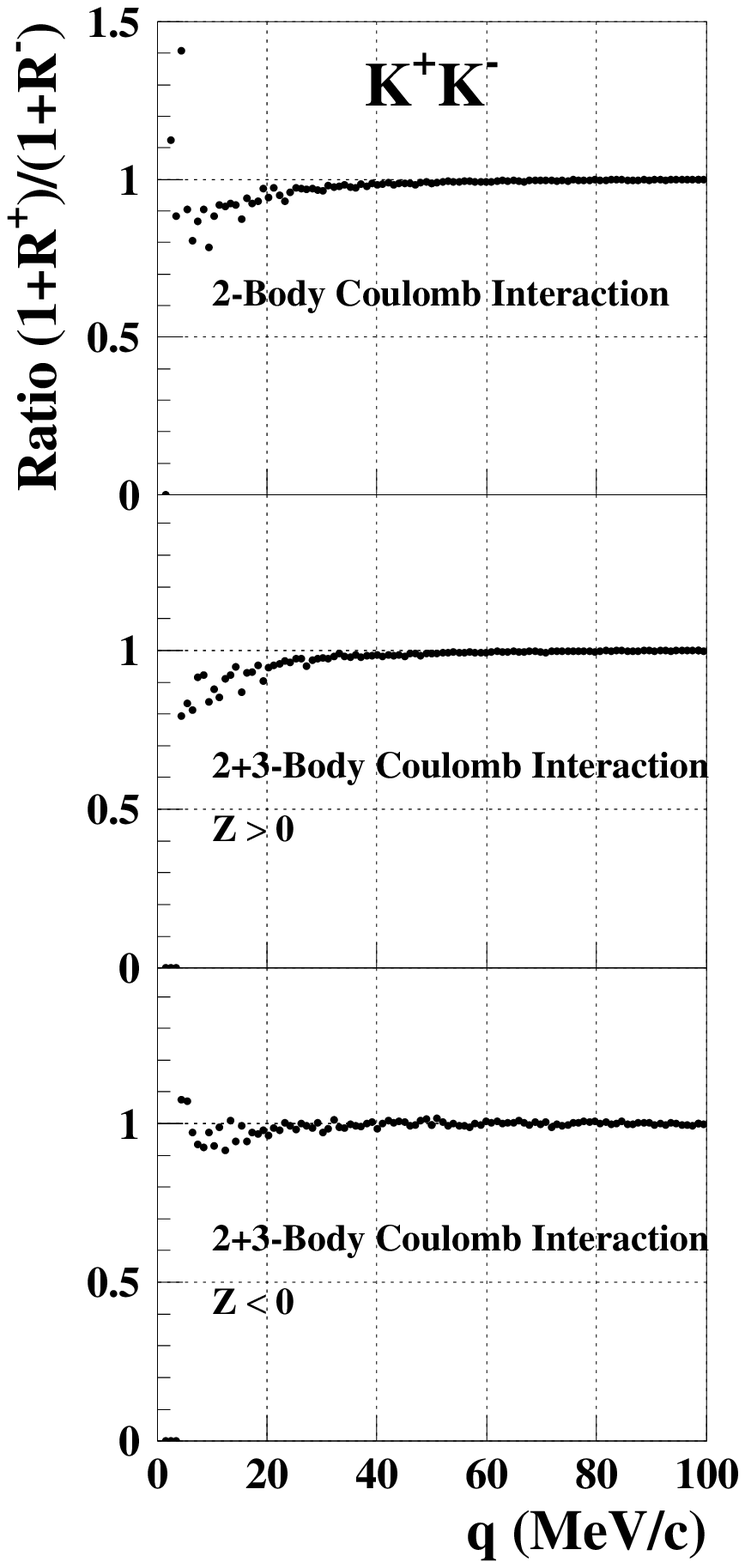}
\vspace*{3cm}
\center{\large \bf Fig.5}}
\newpage
\samepage{
\psfig{figure=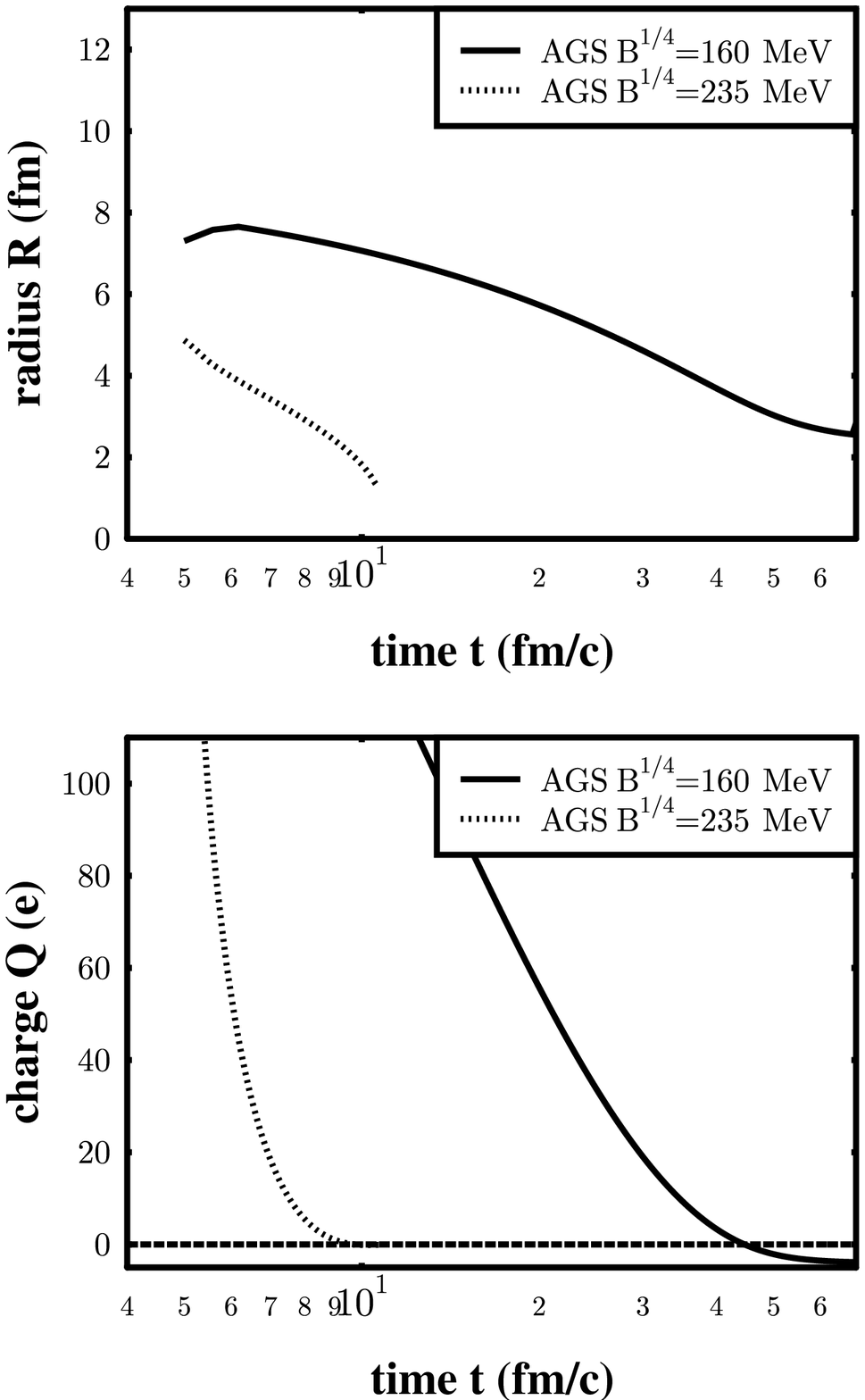}
\vspace*{1cm}
\center{\large \bf Fig.6}}
\newpage
\samepage{
\psfig{figure=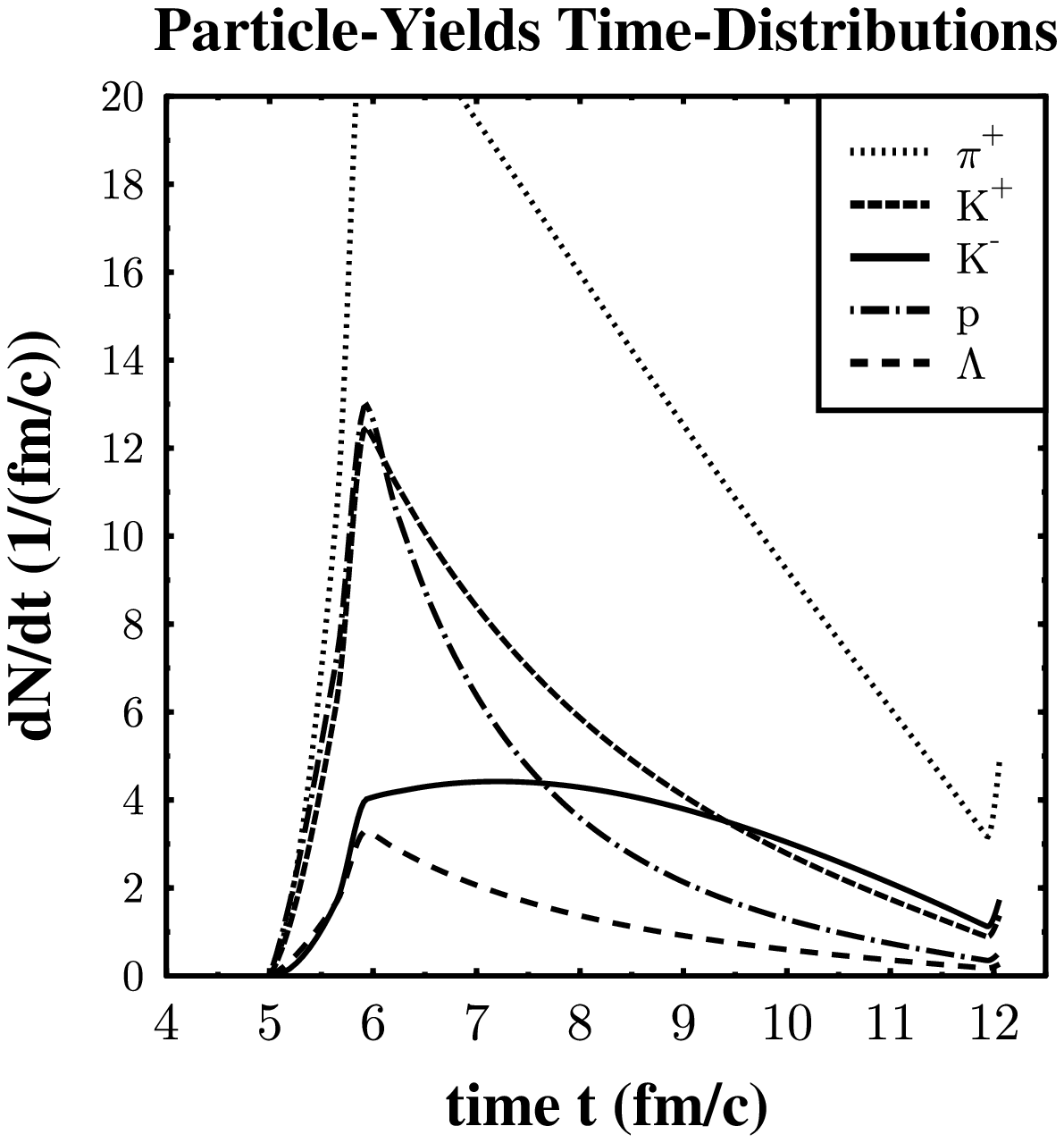}
\vspace*{1cm}
\center{\large \bf Fig.7}}
\newpage
\samepage{\psfig{figure=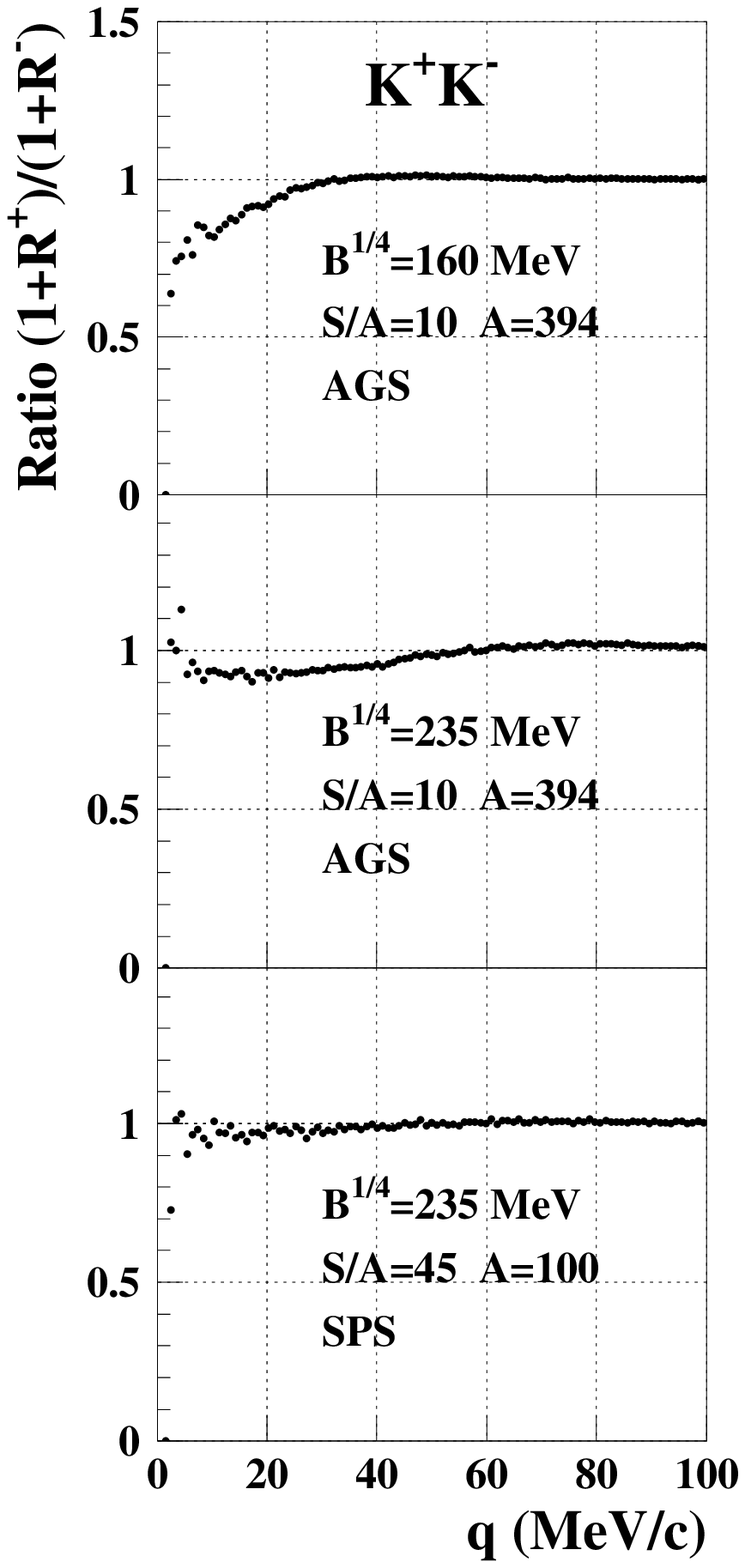}
\vspace*{3cm}
\center{\large \bf Fig.8}}

\enddocument